\documentclass[prx,twocolumn,floatfix,superscriptaddress,longbibliography,nofootinbib]{revtex4-1}

\usepackage{amssymb,amsmath,amstext}
\usepackage{graphicx}
\usepackage{epstopdf}
\usepackage{color}
\usepackage{bm}
\usepackage{appendix}
\usepackage[T1]{fontenc}
\usepackage{bbold}
\usepackage{bbm}
\usepackage{latexsym}
\usepackage[colorlinks=true,citecolor=blue,linkcolor=magenta]{hyperref}

\usepackage[normalem]{ulem}
\usepackage{xcolor}
\usepackage{braket}
\usepackage{soul}

\definecolor{DarkGreen}{RGB}{99,140,98}

\usepackage{dsfont}
\newcommand{\id}{\mathds{1}}
\DeclareMathOperator{\tr}{tr}

\usepackage{qcircuit}

\begin{document}

\title{Scalable Compilation of Resource-Efficient Tapers for High-Confidence Quantum Phase Estimation}

\author{David Raveh}
\email{david.raveh@lanl.gov}
~\affiliation{Los Alamos National Laboratory, Theoretical Division, Los Alamos, NM}
~\affiliation{Department of Physics and Astronomy, Rutgers University, Piscataway, NJ 08854-8019, USA}

\author{Andrew Sornborger}
~\affiliation{Los Alamos National Laboratory, Computing and Artificial Intelligence Division, Los Alamos, NM, USA}

\author{Yi\u{g}it Suba\c{s}\i}
~\affiliation{Los Alamos National Laboratory, Computing and Artificial Intelligence Division, Los Alamos, NM, USA}

\author{Lukasz Cincio}
~\affiliation{Los Alamos National Laboratory, Theoretical Division, Los Alamos, NM}

\begin{abstract}
Quantum phase estimation is a key component of many quantum algorithms. Recently, high-confidence, tapered QPE (tQPE) has been introduced, significantly reducing the number of ancilla qubits necessary for a given failure probability, $\epsilon$. We consider tQPE with $p = l+m$ ancillas, where $l$ represents the number of ancillas necessary to estimate a phase to a given resolution and $m$ represents the number of ``confidence'' qubits added to ensure a desired success probability. With tQPE the failure probability decays double exponentially in $m$, implying that only a few confidence qubits suffice for virtually any problem of interest. A key challenge for tQPE is the construction of the quantum circuit for preparing the taper state on the ancilla register. Here, we develop scalable tensor network methods (previous work was limited to ($l + m = 24$)) giving highly optimized state preparation circuits for optimally concentrated DPSS (Slepian) tapers. We provide an accompanying database of resource-efficient state preparation circuits for $l = 1, \dots, 50$ and $m = 1, \dots, 4$ \cite{raveh2026quantum}. Our circuits are shallow enough for a demonstration on near-term devices for small values of $l$. We run tQPE circuits for a 1-qubit system initialized in the ground state on Quantinuum's H2-2 quantum computer, demonstrating a $3\times$ to $18\times$ reduction in failure probability for $l=1,\ldots,10$ on a noisy device. 
\end{abstract}
\maketitle

\section{Introduction}

The quantum phase estimation (QPE) algorithm uses a quantum computer to estimate the eigenvalue of a unitary operator, $U$, associated with a given input eigenstate, $\ket{\psi}$~\cite{kitaev1995quantum, shor1994algorithms, shor1999polynomial}. This algorithm allows us to estimate the ground state energy of a Hamiltonian with very high accuracy, and may be used as an eigenstate-preparation scheme~\cite{abrams1999quantum}. QPE is versatile and has been used for problems such as estimating molecular ground-state and excited-state energies and solving linear systems of equations~\cite{harrow2009quantum}. It has also been used for quantum principal component analysis~\cite{lloyd2014quantum}, fast quantum Merlin-Arthur amplification~\cite{nagaj2009fast}, and various other tasks~\cite{an2021quantum, chowdhury2018improved, engel2021linear, lloyd2013quantum, rall2020quantum, wright2021automatic, wiebe2012quantum}.

If the system qubits are in an eigenstate, $|\psi\rangle$, of the unitary, $U$,
\begin{equation}
    U|\psi\rangle=e^{2\pi i\theta}|\psi\rangle,\quad \theta\in[0,1)\,,
\end{equation}
then QPE allows us to estimate $\theta$. In QPE, the number of ancilla qubits determines the precision with which $\theta$ is estimated. To achieve a precision $\delta=2^{-(l+1)}$, standard QPE uses $l$ ancilla qubits. Measuring these qubits yields an $l$-bit string encoding the phase estimate, $\hat{\theta}=0.\theta_0\dots\theta_{l-1}$, represented in binary.

Once the desired precision is set, one can then consider the ``confidence'' of the calculation. This quantity is defined as the probability that the calculation achieves an answer within $\delta$ of the true value.
High-confidence QPE refers to variants of QPE that estimate the phase to the desired precision with success probability at least $1-\epsilon$, for a prescribed failure probability, $\epsilon$. Coherent implementations of QPE are required when the phase estimate must remain coherent as part of a larger quantum computation, as in linear-system solvers~\cite{harrow2009quantum} and quantum Metropolis sampling~\cite{temme2011quantum}. In addition, QPE-based ground-state energy estimation from an imperfect initial state also requires a per-run failure probability small compared with the overlap~\cite{berry2025rapid}. High confidence can be achieved in several ways, although the associated resource overhead can vary significantly between methods.
This is particularly important in the fault-tolerant regime, where each independent phase-estimation run may involve a substantial computational cost. Consequently, achieving a prescribed confidence with as few repetitions as possible can be critical.

\begin{figure}
     \centering
\includegraphics[width=1\columnwidth]{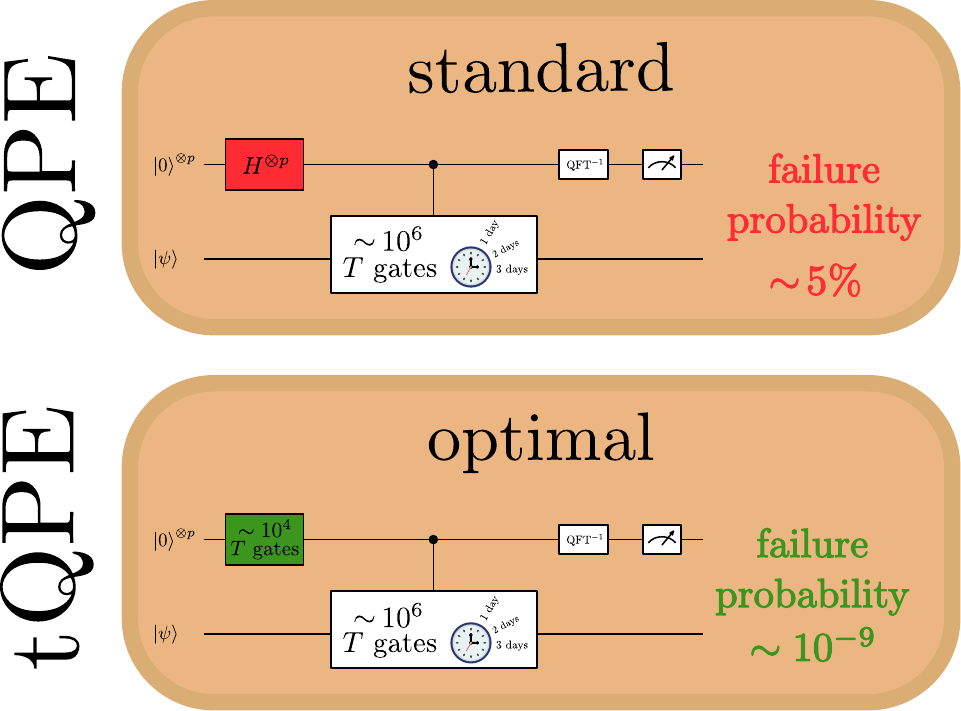}
     \caption{Illustration of tapered QPE in the fault-tolerant regime for a representative megaquop-scale calculation, corresponding here to $\sim 10^6$ $T$ gates. Rather than running QPE with the standard top-hat taper $|+\rangle^{\otimes p}$, we instead prepare the DPSS taper on the ancilla register at a $T$-gate cost of $\sim 10^4$, or approximately $1\%$ of the controlled-evolution cost. The resulting failure probability is exponentially reduced relative to the top-hat taper.}
     \label{fig:tQPE}
\end{figure}

Standard QPE \cite{kitaev1995quantum,abrams1999quantum} suffers from a large failure probability; in the worst case the probability that  $|\hat\theta-\theta|>\delta$ is approximately $60\%$.
One way to mitigate this chance of failure is to add $m$ confidence qubits and run QPE on $p=l+m$ ancilla qubits. The standard approach is to initialize the QPE register in the uniform superposition state $|+\rangle^{\otimes p}$, implementing a Dirac comb, often referred to as a top-hat taper. For the standard QPE algorithm, the number of confidence qubits required to suppress the failure probability to $\epsilon$  is given by \cite{cleve1998quantum}
\begin{equation}
m=O(\log(1/\epsilon)).
\end{equation}

Another method for boosting the success probability is the coherent median method ~\cite{nagaj2009fast}. In this approach, $r=O(\log(1/\epsilon))$ instances of standard QPE, each using $l$ ancilla qubits, are run coherently and in parallel, and the median of the resulting phase estimates is computed coherently. When QPE is used as a coherent subroutine, the median estimate is copied to an additional register and the auxiliary computation, including the individual QPE instances, is subsequently uncomputed. 

Recent work~\cite{berry2024analyzing,patel2024optimal} gave an alternative approach that achieves optimal query complexity $O(\log(1/\epsilon)/\delta)$. The approach considers replacing the top-hat taper $|+\rangle^{\otimes p}$ on the ancilla register with a more general taper $|\phi\rangle$, see Fig. \ref{fig:tQPE}. The optimal taper was shown~\cite{patel2024optimal} to be the discrete prolate spheroidal sequence (DPSS), offering a  \emph{double-exponential} reduction in failure probability versus $m$,
\begin{equation}
m=O(\log\log(1/\epsilon)),
\end{equation}
and an \emph{exponential} improvement relative to \cite{cleve1998quantum} (see Fig.~\ref{fig:MPSfailprob}). 

Both the tQPE approach and coherent median approach achieve the asymptotically optimal query complexity $O(\log(1/\epsilon)/\delta)$, but the coherent median approach has several drawbacks. First, it requires $r\times l$ ancilla qubits to store the $r$ phase estimates. Second, coherently computing the median requires additional circuitry, typically involving a quantum sorting network. Finally, its query complexity has less favorable constants than tQPE for the parameters $l$ and $m$ tested in this work. In contrast, the primary drawback of the tQPE approach is that one needs to construct a circuit which prepares the DPSS taper $|\phi\rangle$.

In \cite{patel2024optimal}, the authors gave an explicit quantum algorithm to construct DPSS tapers that makes use of a quantum Fourier transform (QFT) module. Although accurate, the QFT-based taper preparation circuit in their algorithm can be compressed with tensor network methods. This paper addresses this issue by providing explicit tensor network-derived quantum circuits for DPSS tapers.

The primary goals of this paper are 1) to efficiently compile the DPSS taper to a quantum circuit for a broad range of $l$ and $m$ values that will be of interest to the broader community, 2) to create and make available an open-source database containing these circuits, and 3) to perform initial tests on a near-term device.

In Fig.~\ref{fig:overview}, we give an overview of the tensor network techniques used to obtain circuits for $l=1,\ldots,50$, and $m=1,\ldots,4$; for $m=4$, these circuits achieve $\epsilon<6\times 10^{-14}$ for all $l$. Throughout the paper, $\epsilon$ refers to the worst-case error probability of phase estimates. To attain the worst case in all settings,  we tune the eigenphase to fall in the middle of two successive grid points which depends on $l$ and $m$; see Appendix \ref{sec:model} for details. Our taper circuits are provided in the repository \cite{raveh2026quantum}. Further, we demonstrate the improvement that DPSS offers over the standard $|+\rangle^{\otimes p}$ taper on a noisy quantum computer for small $l,m$ values. We ran tQPE with 10,000 shots on Quantinuum's H2-2 device with various tapers, and while the noise obscures the full improvement potential of a fault-tolerant device, an improvement of $3\times$ to $18\times$ can be seen for $l=1,\ldots,10$.

\section{DPSS Compilation}
\label{sec:DPSScompilation}

\begin{figure}
     \centering
\includegraphics[width=1\columnwidth]{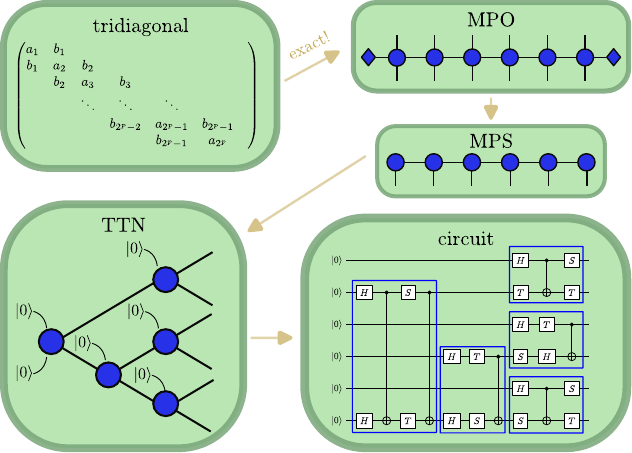}
     \caption{Overview of the approach taken to compile DPSS tapers to quantum circuits. Each step is explained in detail in its own subsection in Sec. \ref{sec:DPSScompilation}. We highlight that while most of these steps are high-accuracy approximations, the MPO representation of the tridiagonal is exact with bond dimension~20, independent of $l$ and $m$.}
     \label{fig:overview}
\end{figure}

The DPSS taper with $l$ precision qubits and $m$ confidence qubits is given by the eigenvector corresponding to the largest eigenvalue of the Slepian frequency concentration matrix \cite{patel2024optimal}
\begin{equation}
    C_{ij}
=
\frac{
\sin\!\left(\pi K(i-j)/N\right)
}{
\pi K(i-j)/N
},
\qquad i,j=0,\ldots,N-1,
\end{equation}
where $N=2^p$, $p=l+m$, and $K=2^m-1$ (rather than $K=2^m$ following the approach in \cite{patel2024optimal}). Note that for $m=0$, the DPSS taper is reduced to the standard top-hat taper $|+\rangle^{\otimes p}$. Slepian showed that the matrix $C_{ij}$ commutes with (and hence has the same eigenvectors as) the tridiagonal matrix defined by \cite{slepian1978prolate}
\begin{align} \label{eq:T}
    T_{ii}&=\left(\frac{N-1}{2}-i\right)^2\cos\left(\frac{\pi K}{N}\right)\\
    T_{i,i+1}&=T_{i+1,i}=\frac{(i+1)(N-1-i)}{2}\,.\nonumber
\end{align}
Further, the DPSS taper is still given by the eigenvector of $T$ with largest eigenvalue. Working with either $C$ or $T$ has advantages  and disadvantages, depending on the context; here, we choose to work with $T$, since $T$ admits an exact and efficient classical representation as a tensor network.

Previous work~\cite{kristjuhan2026quantum} has employed standardized tridiagonal eigensolvers on $T$ for $p=24$ qubits to obtain $2^p$-dimensional eigenvectors representing the DPSS tapers. However, for problems requiring many bits of precision, this approach becomes exponentially challenging. In this section we show how tensor networks allow us to overcome this scalability issue. We will show how we obtain quantum circuits for $l=1,\ldots,50$ and $m=1,\dots,4$; see Fig. \ref{fig:overview} for a diagram of the approach. We first derive an exact matrix product operator (MPO) for the tridiagonal Slepian matrix. The MPO bond dimension required for this construction is $\chi=20$, independent of $l$ and $m$. We then implement the density-matrix renormalization-group (DMRG) algorithm with high-precision on this MPO, obtaining low-$\chi$ matrix product state (MPS) representations of the DPSS tapers. The tridiagonal matrix $T$ is a real symmetric matrix, and hence we solve for a real-valued MPS. We then convert each MPS to a real tree tensor network (TTN) with low bond dimension, as TTNs allow for more general connectivity choices than MPSs, which are naturally restricted to nearest-neighbor connectivity. Finally, we compile each tree tensor in the network to a quantum circuit, searching for optimal qubits $(i,j)$ to place a two-qubit gate and optimizing the gates via environment updates. We take advantage of the fact that the taper is real-valued to consider generic $O(4)$ gates instead of the full $SU(4)$ gate required for general quantum states. The circuits for each tree tensor can then be stitched together to create a quantum circuit which prepares the DPSS taper. The TTN to circuit compilation is based on~\cite{sims2026multivariate}.

\subsection{Tridiagonal $\to$ MPO}

In order to find a circuit for a DPSS taper, we will first need an efficient classical description of the leading eigenvector of the tridiagonal matrix, $T$, defined in Eq.~\eqref{eq:T}. Towards this goal, our first step is to modify our representation of $T$, expressing it in the language of a low-$\chi$ MPO. We now show that $T$ can be exactly represented as an MPO with $\chi=20$, independent of $l$ and $m$. 

To show this, we begin by introducing the $p$-qubit operators
\begin{align}
\hat n &= \sum_{i=0}^{N-1} i\,\ket{i}\!\bra{i},\\
S^+ &= \sum_{i=0}^{N-2} \ket{i+1}\!\bra{i},\qquad
S^- = \sum_{i=1}^{N-1} \ket{i-1}\!\bra{i}.
\end{align}
We then have
\begin{equation} \label{eq:T2}
    T=\cos\left(\frac{\pi K}{N}\right) A^2 + \frac{1}{2}S^+BC + \frac{1}{2}BCS^-,
\end{equation}
where
\begin{align}
    A&=\hat n-\frac{N-1}{2}\nonumber\\
    B&=\hat n+1\nonumber \\
    C&=N-1-\hat n\,.
\end{align}

The operators $X\in\{S^+,S^-,A,B,C\}$ each admit an exact MPO representation with bond dimension $\chi=2$. Explicitly, $X$ is obtained by contracting the auxiliary MPO bond indices according to \cite{schollwock2011density}
\begin{equation}
X_{\mathbf{s},\mathbf{s}'}
=
L_X
W_X^{s_0s'_0}
W_X^{s_1s'_1}
\cdots
W_X^{s_{p-1}s'_{p-1}}
R_X,
\end{equation}
where the products denote contractions over the auxiliary bond indices. For fixed physical indices $s_j,s'_j$, each $W_X^{s_js'_j}$ is a $\chi\times\chi$ matrix acting on the MPO bond space, while $L_X$ and $R_X$ are the left and right boundary vectors. Equivalently, we may regard $W_X$ as a
$\chi\times\chi$ matrix of single-qubit operators, with the two
representations related by
\begin{equation}
\left(W_X^{ss'}\right)_{\alpha\beta}
=
\bra{s}(W_X)_{\alpha\beta}\ket{s'}.
\end{equation}
The MPO tensors below are given in this operator-valued matrix notation. Defining $\sigma^\pm=(\sigma^x\pm i\sigma^y)/2$, the MPO tensors are
\begin{align}
&W_{S^\pm}=
\begin{pmatrix}
\sigma^\pm & 0 \\
\sigma^\mp & \id
\end{pmatrix},
\quad
L_{S^\pm}=
\begin{pmatrix}
0 & \id
\end{pmatrix},
\,
R_{S^\pm}=
\begin{pmatrix}
\id \\
0
\end{pmatrix}
\\
&W_A=
\frac{1}{2}
\begin{pmatrix}
2 & -\sigma^z \\
0 & 4
\end{pmatrix},
\quad
L_A=
\begin{pmatrix}
\id & 0
\end{pmatrix},
\,
R_A=
\begin{pmatrix}
0 \\
\id
\end{pmatrix}
\nonumber\\
&W_B=
\frac{1}{2}
\begin{pmatrix}
2 & \id-\sigma^z \\
0 & 4
\end{pmatrix},
\quad
L_B=
\begin{pmatrix}
\id & 0
\end{pmatrix},
\,
R_B=
\begin{pmatrix}
\id \\
\id
\end{pmatrix}
\nonumber\\
&W_C=
\frac{1}{2}
\begin{pmatrix}
2 & \id+\sigma^z \\
0 & 4
\end{pmatrix},
\quad
L_C=
\begin{pmatrix}
\id & 0
\end{pmatrix},
\,
R_C=
\begin{pmatrix}
0 \\
\id
\end{pmatrix}.
\nonumber
\end{align}
Forming the linear combinations and products of these MPOs required in Eq.~\eqref{eq:T2} gives an explicit MPO representation of the DPSS tridiagonal matrix $T$ with bond dimension $\chi=4+8+8=20$.

\subsection{MPO $\to$ MPS}

\begin{figure}
     \centering
\includegraphics[width=1\columnwidth]{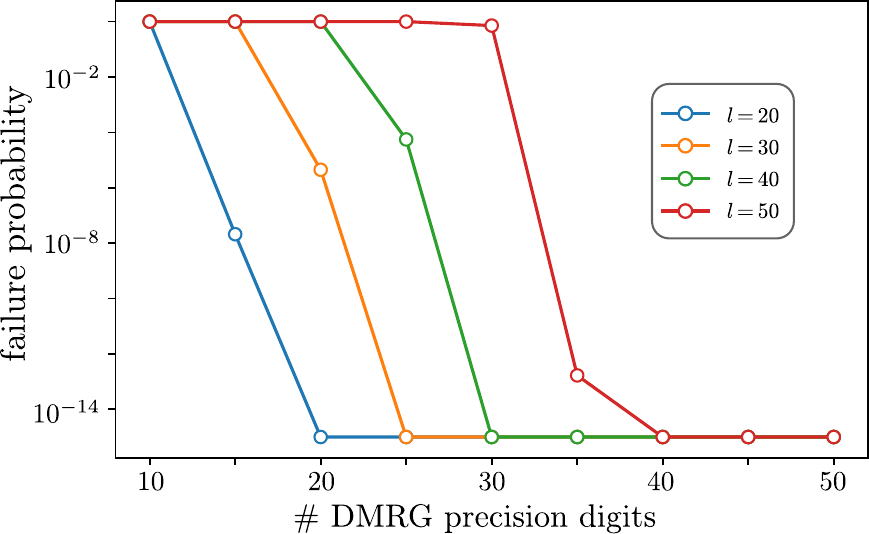}
    \caption{Failure probability $\epsilon$ of the MPSs obtained from DMRG for $m=4$, $\chi=10$, for several values of $l$, as a function of the number of classical precision digits used in the DMRG. While the optimal failure probability is essentially independent of $l$, the number of classical precision digits required to reach it grows approximately linearly with $l$, consistent with the exponentially closing gap, $\Delta\sim4^{-l}$.}

     \label{fig:prec}
\end{figure}

Having such a low-$\chi$ MPO representation of the Hermitian operator, $T$, strongly motivates applying the DMRG algorithm on $-T$ to obtain the ground state of $-T$ represented as an MPS \cite{white1992density,schollwock2011density}. However, this is not so simple: DMRG is an energy-based optimization, and the gap empirically scales exponentially with $l$:
\begin{equation}
    \Delta=\frac{\lambda_0-\lambda_1}{\lambda_0}\sim4^{-l},
\end{equation}
so that the number of (classical) base-10 precision digits required to resolve the gap scales linearly with $l$. This can be tested by implementing DMRG with varying numbers of precision digits. In Fig. \ref{fig:prec} we demonstrate this by computing the failure probability $\epsilon$ of MPSs corresponding to DMRG implemented with various precisions; see Appendix \ref{sec:model} for details. We see that for large $l$, standard machine precision does not suffice to obtain an MPS with the optimal $\epsilon$. 

For each $l=1,\ldots,50$ and $m=1,\dots,4$, we implemented DMRG with $50$ digits of precision and obtained MPS representations of the DPSS tapers  with bond dimensions between 5 and 10. The fact that DPSS tapers admit a low-$\chi$ representation that does not scale with $l$ is something that was observed in \cite{kristjuhan2026quantum}. There, they used exact diagonalization to obtain a statevector for $p=1,\ldots,24$; only then did they convert their vector to an MPS. However, this process naturally incurs an exponential scaling in memory. By rewriting $T$ as a low-$\chi$ MPO, we were able to bypass the exponential scaling, directly deriving our MPSs with DMRG.

Importantly, while the DMRG calculation requires high precision to resolve the exponentially closing gap, the subsequent QPE simulation does not. Once the computed failure probability approaches the numerical precision of the simulation, we no longer attempt to resolve its exact value, since errors at this scale are already negligible for the purposes of this work. We therefore perform the subsequent calculations in double precision and interpret failure probabilities of order $10^{-14}$--$10^{-15}$ as reaching the numerical floor of the simulation.

\subsection{MPS $\to$ TTN}

Any MPS can be converted to a sequential ``staircase'' quantum circuit. Indeed, this is the approach that was taken in~\cite{kristjuhan2026quantum}. However, such circuits have a local structure; we find that including non-local interactions can significantly reduce quantum resources required to prepare the taper. Hence, before compiling our DPSS tapers to a quantum circuit, we choose to convert our MPS to a TTN. Although an MPS can be regarded as a particular tree, here we consider more general TTNs, which allows trees with non-staircase structures and non-local connections. TTNs have a natural interpretation as a real-space renormalization, with each node fusing two ``child'' sites into a single ``parent'' site. We now describe our procedure of converting an MPS to a TTN.

Consider an MPS of $p$ sites, with site $i$ having local  dimension $d_i$. We wish to convert this MPS to a TTN with maximum bond dimension, $\chi$. We begin by computing the $2$-site reduced density matrix, $\rho_{ij}$, for all pairs, $(i,j)$. We then collect and order the eigenvalues $\lambda^{(ij)}_0\geq\lambda^{(ij)}_1\geq\ldots,$ of the reduced density matrices $\rho_{ij}$. We then select the pair that maximizes the retained weight, $\sum_{k=0}^{\chi-1}\lambda_k^{(ij)}$. The selected sites are assigned to a common parent node in the tree, and their joint degrees of freedom are replaced by a single effective site. The replacement is performed by  an isometry, $V \colon \mathbb{C}^\chi\to\mathbb{C}^{d_i}\otimes \mathbb{C}^{d_j}$, which is constructed from the $\chi$ leading eigenvectors of $\rho_{ij}$. The isometry, $V$, can be viewed as a tree tensor with parent index of dimension $\chi$ and child indices with dimensions $d_i$ and $d_j$.

By contracting the child indices of $V^\dagger$ with the MPS sites $i$ and $j$, we obtain a new MPS of $p-1$ sites, with sites $i$ and $j$ fused into a single site with local dimension $\chi$. This procedure can then be repeated, building the TTN by continuing to fuse sites together until there are just two sites left. At that point, the two-site MPS is contracted into one ``top'' tensor (the leftmost tensor shown in Fig.~\ref{fig:overview} and Fig.~\ref{fig:circuit}(a)).

We convert each of the DPSS MPSs to TTNs with bond dimensions between 4 and 7. As an example, a diagram of the TTN for the $l=30,m=3$ DPSS taper can be seen in Fig. \ref{fig:circuit}(a).

\subsection{TTN $\to$ Circuit}

\begin{figure}
     \centering
\includegraphics[width=1\columnwidth]{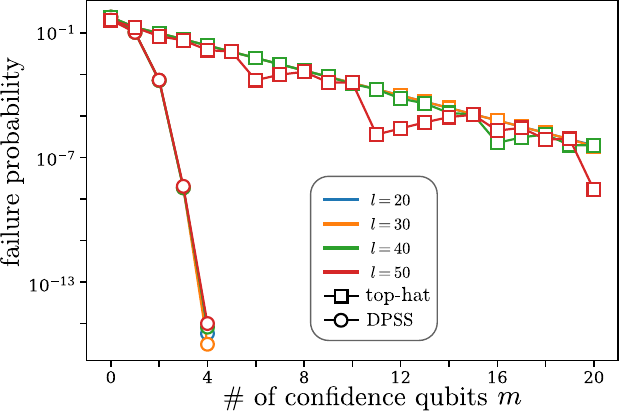}
     \caption{Failure probability of the compiled circuits for various DPSS tapers, compared to the standard top-hat taper $|+\rangle^{\otimes p}$. The top-hat taper gives an exponential reduction in failure probability with $m$ for all values of $l$, while the DPSS taper gives double-exponential reduction. Some numerical instability becomes visible for $p=l+m\gtrsim 55$, where the direct QFT$^{-1}$ simulation involves controlled-phase rotations with angles approaching machine precision. Nevertheless, the qualitative trend remains apparent in this regime.}
     \label{fig:MPSfailprob}
\end{figure}

\begin{figure*}[t]
    \centering
    \includegraphics[width=\textwidth]{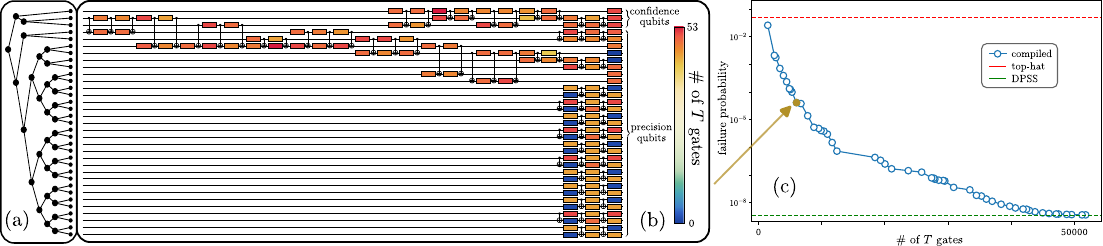}
    \caption{Compilation results for $l=30,\,m=3$. (a) The structure of the TTN used to create the quantum circuits. (b) A circuit diagram with Clifford+$T$ operations for a particular $T$-gate budget; we use a color bar to illustrate the number of $T$ gates in the quantum circuit. As this circuit was generated in the post-compilation stage, where gates contributing the least are deleted, some TTN connections don't appear in the displayed circuit.
    (c) A family of such circuits, showing how failure probability scales with the $T$-gate budget. The family is obtained by testing which gates can be removed at the lowest infidelity cost.}
    \label{fig:circuit}
\end{figure*}

We are now ready to compile the DPSS TTN to a unitary quantum circuit, by compiling each isometric tree element to a quantum circuit consisting of Clifford$+T$ gates. 
Rather than optimizing over the discrete gates $\{H, S,T,\mathrm{CNOT}\}$, we instead compile the isometries to general 2-qubit rotations, and then synthesize these rotations into our discrete gate set. For a generic state, compiling to $SU(4)$ gates is rather natural. However, a generic $SU(4)$ gate consists of $15$ continuous parameters, and will incur a large $T$-gate cost. Since our taper state is real-valued, we can instead choose to compile to $O(4)$ gates, which have just $6$ real parameters. We describe the synthesis of an $O(4)$ circuit into the Clifford$+T$ gate set in Appendix \ref{sec:Tsynth}; for now, we focus on the compilation of our isometries to $O(4)$ gates. 
This has the advantage of allowing us to optimize over continuous parameters instead of discrete gates, and will allow for more general synthesis choices other than Clifford$+T$.

The procedure described here is a modified, low-bond-dimension version of the TTN compilation introduced in~\cite{sims2026multivariate}. 
The isometry to $O(4)$ compilation can be viewed as a multi-state preparation problem. Let the child indices of the isometry, $V$, in question have dimensions $\chi_L$ and $\chi_R$, and the parent have dimension $\chi_P$. Viewing the tensor, $V$, as $\chi_P$ many states, $|\psi_k\rangle$, of dimension $\chi_L\times\chi_R$, we search for a unitary that will convert the basis state, $|k\rangle$, to its corresponding state, $|\psi_k\rangle$, for $k=0,\ldots,\chi_P-1$. We assign $\lceil\log_2\chi_L\rceil$ qubits to the left child index, and similarly for the right, padding the $|\psi_k\rangle$ states with zeros as needed.

For a moment, let us assume we have some ansatz for how the circuit for the tree tensor should look (e.g. a brickwork circuit with a fixed number of layers). With the fixed ansatz, $U=\prod_n u_n$, with $u_n$ being arbitrary $O(4)$ gates, we aim to optimize the $u_n$ gates with the cost function
\begin{equation}
    C=1-\frac{1}{\chi_P}\sum_{k=0}^{\chi_P-1} \Re\, \langle\psi_k|U|k\rangle\,.
\end{equation}
We optimize the gates one at a time: to update a particular $u_n$ gate, we compute its environment, $\mathcal{E}$, so that $C=1-\Re\tr(u_n\mathcal{E})/4$. Since $C$ is linear in $u_n$, the optimal replacement of $u_n$ is found by implementing a singular value decomposition (SVD) on $\mathcal{E}=\mathcal{U}\mathcal{S}\mathcal{V}^\dagger$, given by $u_n\to\mathcal{V}\mathcal{U}^\dagger$. By sweeping repeatedly through the gates in the ansatz, the circuit parameters converge to a near-optimal solution within the chosen ansatz.

However, it is typically not clear what ansatz will work in general. To determine the ansatz, we begin with an empty quantum circuit. Then, for each $(i,j)$ qubit pair, we add an $O(4)$ gate on qubits $i,j$, and optimize the entire circuit; optimizations on the various pairs can be done simultaneously. It should be noted that while the number of pairs to check is scaling quadratically with the number of qubits in the isometry, the isometries have bond dimension $4-7$, and hence have just $4-6$ qubits, making this parallelization feasible. The $(i,j)$ pair that best minimizes the cost function, $C$, is selected. Such a greedy optimization can get trapped in local minima; when this happens, we randomly delete a gate to perturb the circuit, in the form of simulated annealing. Proceeding in this fashion, we continue to build our circuit until the cost function for each tensor is below some threshold. We then stitch together the circuits for each tensor into one global circuit that prepares the target DPSS taper. 

Once the global circuit is constructed, we can further improve it by optimizing the circuit to the original MPS directly (rather than optimizing the pieces separately). This can be done in a similar process, by defining a new cost function, $C_{\text{global}}=1-\Re\,\langle\psi_\text{DMRG}|U_{\text{global}}|0\rangle^{\otimes p}$, we can sweep through all $O(4)$ gates in $U_{\text{global}}$ and optimally replace the gate with an SVD of its environment. Once global optimization is complete, we go through a ``post-compilation'' process of pruning the number of gates in our circuit. For the $O(4)$ gates in our circuit contributing the least entanglement, we remove the gate and recompile the reduced ansatz to the global MPS. We do these computations simultaneously, and remove the $O(4)$ gate that has the smallest impact on the fidelity of the circuit. Repeating this process allows us to obtain a family of circuits approximating the DPSS taper to different compilation errors for varying numbers of $O(4)$ gates.

\begin{figure}
     \centering
\includegraphics[width=1\columnwidth]{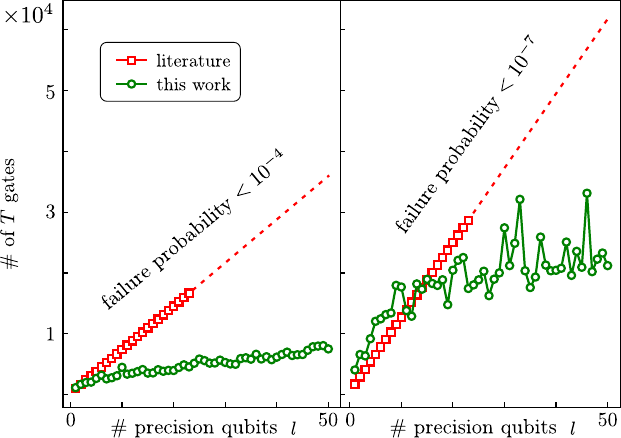}
     \caption{$T$-gate counts for failure probabilities $\epsilon<10^{-4}$ and $\epsilon<10^{-7}$, compared to the results in \cite{kristjuhan2026quantum}, see Appendix \ref{sec:Tsynth} for details. For fixed $m$ and $\epsilon$, the $T$-gate count for \cite{kristjuhan2026quantum} scales linearly in $l$, while our approach appears to scale logarithmically. For $\epsilon=10^{-4}$ and $\epsilon=10^{-7}$, our approach gives significant $T$-count reductions over that approach. The dashed red line indicates extrapolated results, as the method in \cite{kristjuhan2026quantum} is restricted to $p\leq24$.}
     \label{fig:Tgates}
\end{figure}

We compiled the DPSS tapers with $l=1,\ldots,50$ and $m=1,\ldots,4$, creating a family  of $O(4)$ circuits for each $(l,m)$, and synthesized each circuit to Clifford+$T$ as described in Appendix \ref{sec:Tsynth}. These circuits are available in the repository \cite{raveh2026quantum}. Our compilation results are shown in Figs.~\ref{fig:MPSfailprob}--\ref{fig:Tgates}. 
Fig.~\ref{fig:MPSfailprob} shows the effect of adding $m$ confidence qubits on the failure probability for various $l$. The standard top-hat taper $|+\rangle^{\otimes (l+m)}$ offers exponential reduction in failure probability, but the DPSS taper expressed as a compiled quantum circuit offers \emph{double-exponential} reduction. In Fig.~\ref{fig:circuit}(c), we show the family of circuits with varying $T$-gate counts and failure probabilities, $\epsilon$. In Fig.~\ref{fig:circuit}(b), we show a circuit diagram of a particular circuit in the family, with a color bar that indicates the number of $T$ gates in the single-qubit gates. The structure of the circuit appears to indicate that the most significant qubits are largely in a product state, with most of the gates concentrated on generating entanglement on the least significant qubits. This structure is not at all obvious, and highlights the value of building an ansatz adaptively to a particular state. Fig.~\ref{fig:Tgates} shows a comparison of our compilation results with the results of  \cite{kristjuhan2026quantum}; see Appendix \ref{sec:Tsynth} for details. The comparison indicates that our approach gives a substantial improvement over the approach in \cite{kristjuhan2026quantum} for large $l$. It should be noted that Ref.~\cite{kristjuhan2026quantum} achieved quantum circuits for $p\leq24$ down to $R_z$ gates, by converting their statevectors into an MPS; hence for $p>24$ the counts they report are extrapolations, as they did not derive MPSs for those cases. Further, their circuit construction requires $\chi\leq4$; that bond dimension is insufficient to reach precision beyond $10^{-9}$. Hence, their approach would fail to achieve the $m=4$ result of $\epsilon<6\times10^{-14}$ even for moderate values of $l$. 

The family of Clifford+$T$ circuits obtained from our approach, allowing a tradeoff between $\epsilon$ and the $T$-gate budget (see Fig.~\ref{fig:circuit}c), is available in the repository \cite{raveh2026quantum} for each $(l,m)$. We recognize that Clifford+$T$ is not the only gate set relevant to fault-tolerant quantum computing, as the gate set is both hardware specific and error-correction-code specific. Hence, we also include the family of $O(4)$ circuits for each $(l,m)$ in the repository. This way, users can download our $O(4)$ circuits, and synthesize them to their preferred gate set.

\section{Hardware Implementation}

Running QPE is typically considered a task for fault-tolerant quantum computers. However, implementations on noisy devices available today are possible for small one- or two-qubit systems with a moderate ancilla register \cite{tranter2025high}. The fact that the DPSS tapers can be compiled to few $O(4)$ gates motivates demonstrating the advantage of the DPSS taper over the top-hat taper even on a noisy device. Noise makes it exceedingly difficult to observe an improvement, since just a few entangling gates contribute enough noise to obscure the improvement the DPSS taper would give in a noiseless setting. Hence, we consider certain ``hardware-informed'' modifications to our quantum circuits that will maximize the improvement shown on the hardware. In this section we will show how we modified our tQPE circuits for a 1-qubit system to run on Quantinuum's H2-2 device, and then show the results from the hardware.

\subsection{Hardware-informed compilation}

Quantinuum's H2-2 device uses trapped $^{171}\mathrm{Yb}^{+}$ ions as qubits, has a 2-qubit gate fidelity of 99.917\%, and boasts all-to-all connectivity. The entangling gate is $ZZ(\vartheta)=e^{-iZZ\vartheta/2}$, and the noise increases linearly with $\vartheta$. Let us consider the extreme $l=1$ case: here the global tQPE circuit will have just a handful of entangling gates. Still, the noise level of a few $ZZ$ gates is enough to cause failure errors at a rate $\epsilon>10^{-3}$. Hence, the noise could completely obscure the improvement an $m=3$ taper has to offer. The improvement the $m=1$ taper offers is quite small; hence, we restrict ourselves to running $m=2$ tapers for small $l$, observing the impact of noise as $l$ increases.

\begin{figure}[!t]
     \centering
\includegraphics[width=1\columnwidth]{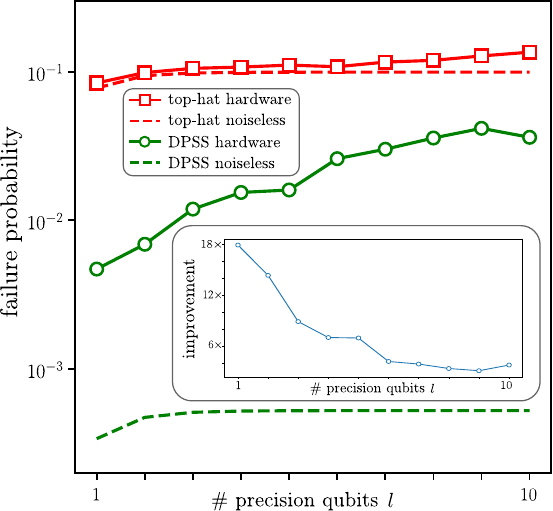}
     \caption{Failure probability for top-hat and DPSS tapers with $m=2$ as a function of the number of precision qubits $l$. Hardware results were obtained on Quantinuum's H2-2 device using 10,000 shots per circuit, while the corresponding noiseless results are shown for comparison. The DPSS taper reduces the failure probability for all $l=1,\ldots,10$, with a $3\times$--$18\times$ improvement on hardware. The inset shows the ratio of the top-hat to DPSS hardware failure probabilities. Although device noise increasingly obscures the ideal DPSS advantage as $l$ grows, the improvement remains visible throughout the range studied. Note that for each $l$ the eigenphase was shifted to lie at the midpoint between two grid points, so the failure probability isn't fluctuating with each $l$.}
     \label{fig:device}
\end{figure}

The circuits generated in Sec.  \ref{sec:DPSScompilation} were compiled to $O(4)$ gates, and standard \texttt{Qiskit} tools can be used to decompose these gates to $ZZ$ gates and 1-qubit gates. However, this tends to overestimate the number of $ZZ$ gates needed to prepare the state. Hence, rather than compiling the TTN isometries to general $O(4)$ gates, we instead compile them directly to $ZZ(\vartheta)$ and $SU(2)$ gates. We took these circuits through a post-compilation stage as well, searching for the optimal $ZZ$ gates to remove from the circuit, giving us a family of circuits with various $ZZ$-gate counts for each taper. The optimal choice on a noisy device is \emph{not} necessarily the circuit with the highest fidelity to the DPSS taper; instead we should balance the fidelity with the noise coming from the number of $ZZ$ gates. This balance can be found using a sufficiently accurate characterization of the target device’s noise, as we do here.

For various $l$, we ran QPE on Quantinuum's H2-2E noisy emulator for a family of circuits with varying numbers of $ZZ$ gates for 10,000 shots each. For each $l$, the circuit which gave the smallest $\epsilon$ was selected to run on the actual hardware. 

We also considered implementing a truncated QFT$^{-1}$ to reduce the number of gates in our QPE circuit. The basic idea is to omit controlled-phase gates in the QFT$^{-1}$ which act on qubits $i,j$ with $|i-j|>k$, as the angle in the phase gate decays exponentially with $|i-j|$. However, our simulations showed that truncated QFT$^{-1}$ will only give substantial improvements for large $l$, so we chose not to truncate QFT$^{-1}$ for our noisy runs with small $l$.

\subsection{Hardware results}

\begin{table}[t]
\centering
\caption{The number of $ZZ$ gates for each component of the QPE for the $m=2$ hardware runs. We see that the noise is dominated by the QFT$^{-1}$, where the number of $ZZ$ gates scales quadratically.}
\label{tab:zzgates}
\begin{tabular}{c|ccc}
\hline
$l$ & DPSS preparation & Controlled evolution & QFT$^{-1}$ \\
\hline
1  & 3  & 3  & 3  \\
2  & 6  & 4  & 6  \\
3  & 8  & 5  & 10 \\
4  & 4  & 6  & 15 \\
5  & 9  & 7  & 21 \\
6  & 6  & 8  & 28 \\
7  & 11 & 9  & 36 \\
8  & 8  & 10 & 45 \\
9  & 10 & 11 & 55 \\
10 & 9  & 12 & 66 \\
\hline
\end{tabular}
\end{table}

We ran tQPE on the H2-2 quantum computer for $l=1,\ldots,10$ and $m=2$ with both top-hat and DPSS tapers, with 10,000 shots per experiment. Table \ref{tab:zzgates} gives the $ZZ$ gate counts of the selected circuits for each component of the QPE; the QFT$^{-1}$ dominates the taper overhead, with a gate count that scales quadratically with $l$. In Fig.~\ref{fig:device} we compare the failure probabilities of the top-hat and DPSS tapers. In all cases the DPSS taper reduces the failure probability on the hardware, giving improvements of $3\times$ to $18\times$ even on a noisy device.

It is worth highlighting that noise---coming predominantly from the QFT$^{-1}$---still obscures most of the improvement the DPSS taper offers. Indeed, for $m=2$, noiseless circuits give more than $100\times$ improvement for all $l$. We thus conclude that while tQPE gives an improvement even on near-term hardware, its full impact will only be seen in the fault-tolerant regime. 

\section{Discussion}

The main result of this work is a scalable route from optimal DPSS tapers to resource-efficient quantum circuits. By exploiting the tensor-network structure of the Slepian problem, we are able to construct and compile tapers far beyond the system sizes accessible to direct state-vector methods, reaching $l=50$ precision qubits and $m=1,\ldots,4$ confidence qubits. For $m=3$--$4$, the resulting failure probabilities are small enough for most applications. Importantly, the additional cost of preparing the taper remains modest compared with the controlled time evolution expected in fault-tolerant applications.

To make these results directly usable, we provide a repository \cite{raveh2026quantum} of compiled taper-preparation circuits over the full range of $l$ and $m$ considered here. We include both Clifford+$T$ circuits and the underlying $O(4)$ circuits. The latter are particularly useful because they separate the tensor-network compilation from the choice of fault-tolerant gate set: the same compiled $O(4)$ circuit can subsequently be synthesized according to the native gates and error-correction architecture of a particular quantum computer. This makes the repository of circuits more general than a collection of exclusively Clifford+$T$ circuits.

The principal use case for these tapers is fault-tolerant quantum phase estimation, where the cost of a failed phase-estimation run can be substantial. However, our hardware experiments show that the advantage of tQPE is already visible on present-day devices: for $m=2$, the DPSS taper reduced the measured failure probability by factors of $3$--$18$ on Quantinuum H2-2. Nevertheless, noise in the remainder of the QPE circuit obscures much of the much larger advantage predicted in the noiseless case. We therefore view the hardware results primarily as a demonstration that the taper advantage survives realistic noise, while the main impact of the method is expected in the fault-tolerant regime. There, the combination of very small failure probability, modest taper-preparation overhead, and reusable compiled circuits makes DPSS tapering a practical component for high-confidence coherent phase estimation.

\section*{Acknowledgments}

D.R. and Y.S. were supported by the Quantum Science Center (QSC), a National Quantum Information Science Research Center of the U.S. Department of Energy (DOE). A.T.S. acknowledges support from the ASC Beyond Moore's Law project. L.C. was supported by the Laboratory Directed Research and Development (LDRD) program of Los Alamos National Laboratory (LANL) under project number 20260043DR.

\appendix

\section{Model and tQPE simulation}
\label{sec:model}

To test tQPE, we consider an energy estimation problem for a $1$-qubit system. Similar to \cite{tranter2025high}, we choose the H$_2$ molecule at equilibrium atomic separation $0.7414\,\text{\AA}$, given by the Hamiltonian
\begin{align}
    H=&-0.328717028208313\,\id\nonumber\\&-0.181288808211496\,\sigma_x\nonumber\\
    &-0.787967358877028\,\sigma_z\,.
    \label{hamiltonian}
\end{align}
To obtain this Hamiltonian, we used the \texttt{PySCF} quantum chemistry package to obtain the second-quantized Hamiltonian restricted to the STO-3G basis. We then used the \texttt{Qiskit Nature} package to leverage the $\mathbb{Z}_2$ particle-number and spin parities to reduce the system to a 1-qubit Hamiltonian $H=H_{\text{elec}}+H_{\text{nuc}}$, where $H_{\text{nuc}}$ is a constant energy shift corresponding to nuclear repulsion.

Throughout the paper, when we compute the failure probability, we run tQPE on the unitary $U=e^{-iHt}$ with $t=1$, initializing the system qubit in the ground state of $H$. As $l$ and $m$ change, the relative position of the ground state eigenphase of $H$ will change, due to the changing bin size of the tQPE grid. Hence, for fixed $l,m$, we shift $H\to H+c\id$, with $c$ chosen so that the eigenphase lies exactly in the middle of two grid points; this is the provable worst case for the top-hat taper and extremely close to the worst case of the DPSS taper. We applied the controlled $U^{2^i}$ gates, Hadamards, and controlled phase gates on the original MPS (including the system qubit), obtaining an MPS of $p+1$ qubits corresponding to the results of the tQPE algorithm for various taper choices. To compute the failure probability of the resulting MPS, it suffices to look at the squared overlap of the MPS with basis states corresponding to bins within $\delta$ from the exact value, tracing out the system qubit. The number of such bins is $2^m$, independent of $l$.

\section{$T$-gate synthesis and comparison to \cite{kristjuhan2026quantum}}
\label{sec:Tsynth}

\begin{figure}
     \centering
\includegraphics[width=1\columnwidth]{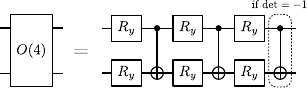}
     \caption{The decomposition of an $O(4)$ gate into six $R_y$ gates \cite{wei2012decomposition}, each of which we efficiently synthesize into Clifford+$T$ gates \cite{ross2014optimal}. We note that the extra $\mathrm{CNOT}$ that appears for an $O(4)$ unitary $U$ with $\det U=-1$ can be placed on the left or on the right.}
     \label{fig:O4synth}
\end{figure}

\begin{figure}
     \centering
\includegraphics[width=1\columnwidth]{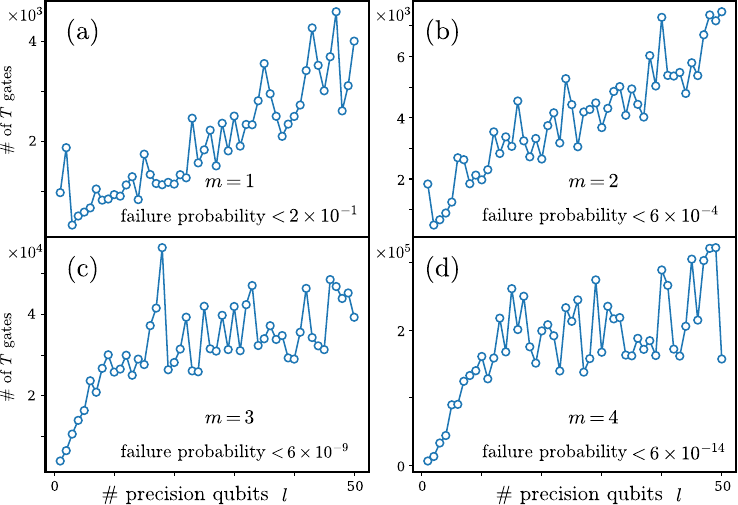}
    \caption{Number of $T$ gates required to achieve the indicated threshold failure probability for $l=1,\ldots,50$, shown for $m=1,\ldots,4$. The $T$-gate count increases overall with the number of precision qubits and grows rapidly with $m$, from $O(10^3)$ gates for $m=1,2$ to $O(10^5)$ for $m=4$. Point-to-point fluctuations reflect the discrete circuit synthesis and optimization used to satisfy the target failure probability. The threshold for each $m$ is shown in its respective panel.}
     \label{fig:totalcosts}
 \end{figure}

The approach in Sec. \ref{sec:DPSScompilation} describes how we compiled the DPSS tapers into quantum circuits consisting of general $O(4)$ gates. To synthesize such a circuit into Clifford$+T$, we decompose each $O(4)$ gate into six $R_y$ gates \cite{wei2012decomposition}, see Fig.~\ref{fig:O4synth}. Note that an $O(4)$ unitary, $U$, has $\det U=\pm1$. In the case where $\det U=-1$, there is an extra CNOT; but we are free to choose which side to place the extra $\mathrm{CNOT}$. Some $R_y$ gates can be trivially combined, e.g., two neighboring $O(4)$ gates with $\det U = +1$ which share a qubit. For the gates with $\det U=-1$, we choose the side to place the extra $\mathrm{CNOT}$ so as to maximize the number of $R_y$ combinations; then we convert $R_y$ gates to $R_z$ gates using $R_y(\theta)=S H R_z(\theta) H S^\dagger$. With a Clifford+$R_z$ quantum circuit, we efficiently synthesize each $R_z$ gate into Clifford+$T$ gates using the approach in ~\cite{ross2014optimal}. In Fig.~\ref{fig:totalcosts} we show the total number of $T$ gates required for each $(l,m)$ when compiling to high accuracy; we see that we were able to compile the $m=4$ tapers with enough accuracy to achieve $\epsilon<6\times10^{-14}$ for all $l=1,\ldots,50$. The plots show large fluctuations in the number of required $T$ gates; we believe this is in part due to the greedy optimization approach. 

We now explain how we compare the $T$-gate counts in our approach with the approach in \cite{kristjuhan2026quantum}. Importantly, while \cite{kristjuhan2026quantum} constructed quantum circuits consisting of Clifford+$R_z$ gates, these $R_z$ rotations were not explicitly synthesized into Clifford+$T$. Instead, they cited the scaling of the required $T$-gate count for a given synthesis precision, where the scaling they used assumes mixed fallback techniques, a non-deterministic approach which requires the use of an ancilla \cite{kliuchnikov2023shorter}. However, since we are aiming to achieve deterministic (non-mixed) unitary synthesis without ancillas, to make the comparison we assume their $T$-gate count to scale as $3.02\log_2(1/\epsilon_{\mathrm{synth}})+1.77$, corresponding to a deterministic unitary synthesis \cite{ross2014optimal,kliuchnikov2023shorter}. We expect that mixed fallback techniques could be utilized for our $R_z$ gates as well, if desired. Following Ref.~\cite{kristjuhan2026quantum}, we set the single-rotation synthesis tolerance equal to the target tQPE failure probability, $\epsilon_{\mathrm{synth}}=\epsilon$. Note that this choice of synthesis error is somewhat arbitrary, but it allows us to do a fair comparison between the two approaches.

\bibliography{quantum}

\end{document}